\documentclass[reprint,amsmath,amssymb,prb,aps,citeautoscript,a4paper,showpacs,showkeys]{revtex4-1}

\usepackage{graphicx}
\usepackage{amsmath}
\usepackage{multirow}
\usepackage{float}
\usepackage{dcolumn}
\usepackage{pdfpages}
\usepackage{hyperref}
\hypersetup{
  final,
  pdfborder={0 0 0},
  pdftitle={Solid-state amorphization of Cu nanolayers embedded in
    a Cu64Zr36 glass},
  pdfauthor={T. Brink, D. Sopu, K. Albe},
  colorlinks=true,
  urlcolor=red,
  linkcolor=black,
  citecolor=blue
}

\newcommand{\uglass}{\ensuremath{u_\text{am}}}
\newcommand{\ufcc}{\ensuremath{u_\text{fcc}}}
\newcommand{\UBMG}{\ensuremath{U_\text{MG}}}

\newcommand{\gammacg}{\ensuremath{\gamma_\text{c--g}}}
\newcommand{\gammagg}{\ensuremath{\gamma_\text{g--g}}}
\newcommand{\cuzr}{Cu$_{64}$Zr$_{36}$}

\makeatletter
\newcommand*{\balancecolsandclearpage}{%
  \close@column@grid
  \cleardoublepage
  \twocolumngrid
}
\makeatother

\begin{document}

\title{Solid-state amorphization of Cu nanolayers embedded in a
  Cu$_\mathbf{64}$Zr$_\mathbf{36}$ glass}

\author{Tobias Brink}
\email{brink@mm.tu-darmstadt.de}
\affiliation{Fachgebiet Materialmodellierung, Institut f{\"u}r
  Materialwissenschaft, Technische Universit\"at Darmstadt,
  Jovanka-Bontschits-Str.~2, D-64287 Darmstadt, Germany}

\author{Daniel \c{S}opu}
\altaffiliation[Present address: ]{IFW Dresden, P.O.\ Box 270116,
  D-01171 Dresden, Germany}
\affiliation{Fachgebiet Materialmodellierung, Institut f{\"u}r
  Materialwissenschaft, Technische Universit\"at Darmstadt,
  Jovanka-Bontschits-Str.~2, D-64287 Darmstadt, Germany}

\author{Karsten~Albe}
\affiliation{Fachgebiet Materialmodellierung, Institut f{\"u}r
  Materialwissenschaft, Technische Universit\"at Darmstadt,
  Jovanka-Bontschits-Str.~2, D-64287 Darmstadt, Germany}

\date{May 6, 2015}

\begin{abstract}
  Solid-state amorphization of crystalline copper nanolayers embedded
  in a Cu$_{64}$Zr$_{36}$ metallic glass is studied by molecular
  dynamics simulations for different orientations of the crystalline
  layer.  We show that solid-state amorphization is driven by a
  reduction of interface energy, which compensates the bulk excess
  energy of the amorphous nanolayer with respect to the crystalline
  phase up to a critical layer thickness.  A simple thermodynamic
  model is derived, which describes the simulation results in terms of
  orientation dependent interface energies.  Detailed analysis reveals
  the structure of the amorphous nanolayer and allows a comparison to
  a quenched copper melt, providing further insights into the origin
  of excess and interface energy.
  \vspace{0.25\baselineskip}\\
  {
    \noindent
    \footnotesize
    Published in:\\
    \href{http://link.aps.org/doi/10.1103/PhysRevB.91.184103}
         {T.~Brink \textit{et al.},
          Phys.\ Rev.\ B \textbf{91}, 184103 (2015)}
    \hfill
    DOI: \href{http://dx.doi.org/10.1103/PhysRevB.91.184103}
              {10.1103/PhysRevB.91.184103}\\%
    \copyright{} 2015 American Physical Society.
  }
\end{abstract}






\raggedbottom

\maketitle

\section{Introduction}

Metals show a strong tendency to form crystalline phases and only
appear in an amorphous state under certain conditions.
Using far-from-e\-qui\-lib\-ri\-um processes, metals can be kinetically
trapped in a metastable state.  This includes bulk metallic glasses
(BMG), which are highly alloyed metallic systems quenched from the
melt.  The glass formation is supported by rapid quenching and the
size difference of the component atoms
\cite{Klement1960,Inoue1989,Inoue1990}.
Furthermore, a crystalline metal sample can be forced into a
disordered state by ion irradiation.  The disorder is introduced by
high-energy impacts of ions, which disturb the ordered lattice due to
local melt--quench processes \cite{Averback1997}.
Thin films produced with high deposition rates can also be amorphous
\cite{kazmerski80}.  In this case, the amorphous state is metastable
and only induced due to the high growth rate in the deposition
process, while in equilibrium thin metal films on a variety of
substrates usually are crystalline.  Examples include iron on
amorphous carbon substrates \cite{Yoshida1972}; Cu, Ag, Al, Au, and Ni
on sapphire substrates \cite{Bialas1994,Moller1991,Dehm1995}; and Ni
on tungsten substrates \cite{Koziol1987} among many others.
Amorphization is not limited to far-from-equilibrium processes, but
can also happen for purely energetic reasons.  High-angle, high-energy
grain boundaries in some polycrystalline metal systems exhibit an
amorphous structure due to the misorientation of the neighboring
crystal lattices \cite{Keblinski1999}.
Molecular dynamics (MD) computer simulations on single-component
\cite{Wolf1995} and binary alloy \cite{Zheng2007} Lennard-Jones
systems identify a nanocrystalline instability.  Nanocrystalline
materials with grain sizes smaller than a critical value become
unstable and collapse completely, leaving behind an amorphous metal,
i.e., the grain boundary phase.  Similarly, metallic nanoparticles
below a critical size can also occur in an ``amorphous'' phase driven
by the reduction of surface energy \cite{Baletto2005}.

In heterogeneous interfaces between crystalline metals, amorphous
interphases were also found \cite{Schwarz1983}.  These interface
interphases are thermodynamically stable and result from the
misorientation and lattice mismatch between the adjacent crystallites
\cite{Benedictus1996,Benedictus1998}.  This effect is called
solid-state amorphization (SSA) and has recently also been discussed
in the framework of complexion formation \cite{Cantwell2014}.  Similar
to the formation of interface interphases, a thin metallic film
embedded in a different crystal phase can transform into an amorphous
state if the thickness is below a critical value
\cite{Landes1991,Handschuh1993,Herr2000}.
While energetically driven amorphization of a thin crystalline layer
due to size mis\-match and misorientation to the abutting crystalline
phases is a well known phenomenon, the amorphization of a thin
elemental metal layer embedded in an amorphous matrix appears to be
less likely, since the driving force should be significantly smaller.

Ghafari \textit{et al.}, however, showed recently that iron
nano\-layers embedded in an amorphous glass
(Co$_{75}$Fe$_{12}$B$_{13}$) can become amorphous, if the thickness is
five monolayers (ML) or less, while at six or more monolayers a
crystalline phase is observed \cite{Ghafari2012}. Whether or not this
is a kinetic effect due the deposition conditions or an energetically
driven phenomenon, which might depend on the lattice orientation, is
the problem of interest in this study.

\begin{figure*}
  \centering
  \includegraphics{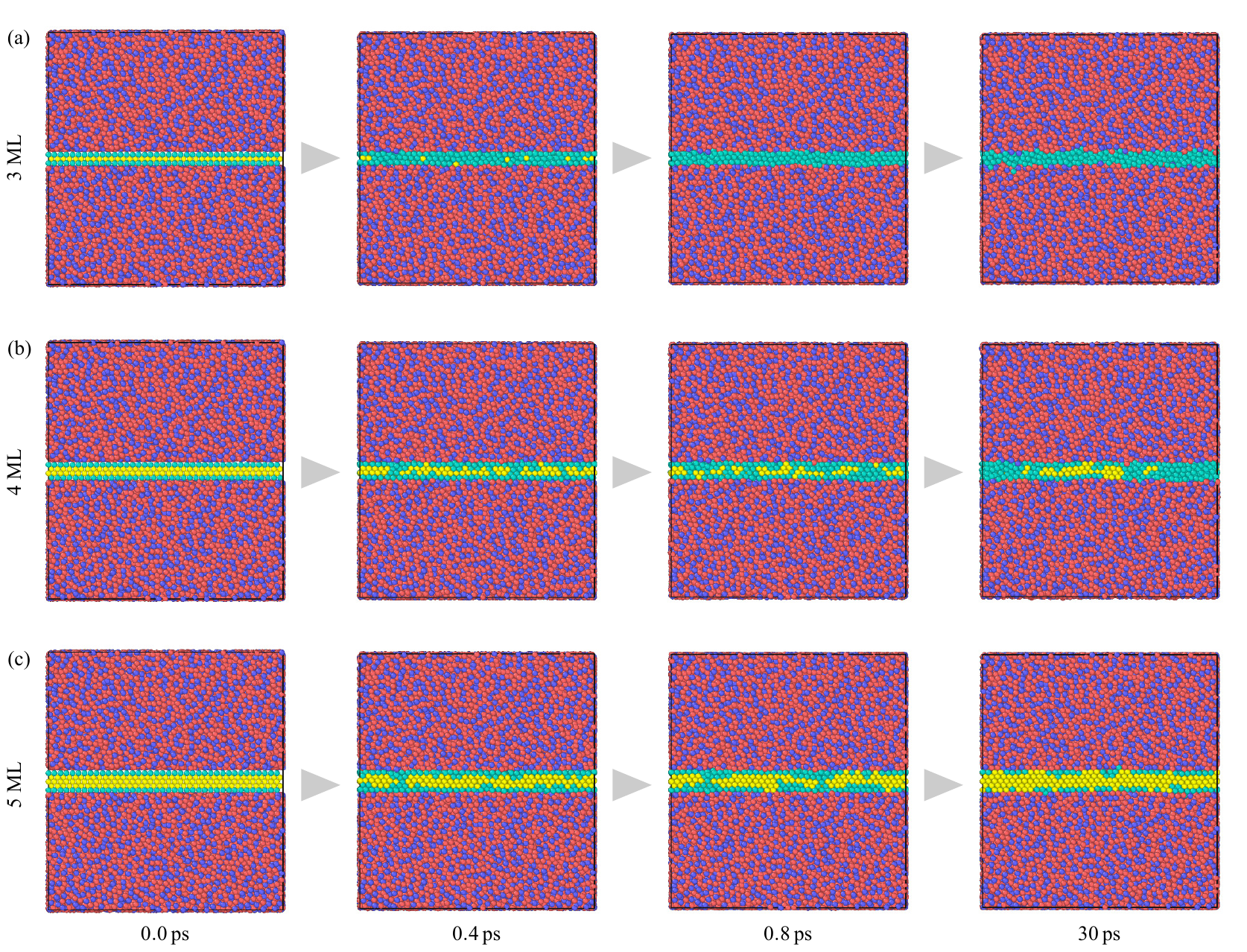}
  \caption{Time evolution of the composite \cuzr{}--Cu systems with
    different nanolayer thickness.  Exemplarily, we show a system in
    which a copper nanolayer with fcc (100) surface was inserted.  The
    top row (a) shows the evolution of the system with a copper
    nanolayer thickness of three monolayers.  The nanolayer in this
    system amorphizes almost immediately.  The middle row (b) depicts
    a system with four monolayers of copper, which stay partly
    crystalline. The bottom row (c) contains snapshots of a system
    where the nanolayer has a thickness of five monolayers.  Here, the
    nanolayer stays crystalline.  This simulation used the Mendelev
    potential.  Copper atoms are shown in red, zirconium atoms in
    blue.  Copper atoms that belong to the inserted layer are green,
    except for those in fcc configuration, which are shown in yellow.}
  \label{fig:seq-sandwich}
\end{figure*}

In order to address this, we conducted MD simulations of Cu nanolayers
of different orientation embedded in a \cuzr{} matrix and investigated
the driving force behind the amorphization.  Since there is no
adequate iron alloy potential which also correctly models an alloy
that forms a metallic glass (MG), we instead used a system based on Cu
and Zr.  To investigate the thermodynamic stability of the amorphous
layer, we start from crystalline nanolayers of varying thickness and
check if there is a phase transition to the amorphous state at a
critical thickness.  We then develop a simple thermodynamic model
based on the assumption that any amorphization in this setup must be
energetic in nature and check it against our simulation results.

\section{Simulation methods and sample preparation}
\label{sec:methods}

\subsection{Simulation method}

\vspace{-3pt}
\noindent
\begin{minipage}{\linewidth}
\setlength\parindent{10pt}
We conducted MD simulations using \textsc{lammps} \cite{Plimpton1995}.
The potential energy was modelled by a Finnis-Sinclair type potential
by Mendelev \textit{et al.}\ \cite{Mendelev2009}, which was mostly
fitted to the glassy state, and another potential by Ward \textit{et
  al.} This second potential consists of elemental potentials by Zhou
\textit{et al.}\ \cite{Zhou2004} that were combined into an alloy
potential by fitting to several intermetallic phases using the Ward
method \cite{Ward2012preprint}.  To obtain independent confirmation of
our simulation results, we ran simulations with both potentials.  The
time-step length was set to $2\,\mathrm{fs}$.  In all simulations we
employed periodic boundary conditions to obtain a surface-free system.
\end{minipage}

\begin{figure*}
  \centering
  \includegraphics{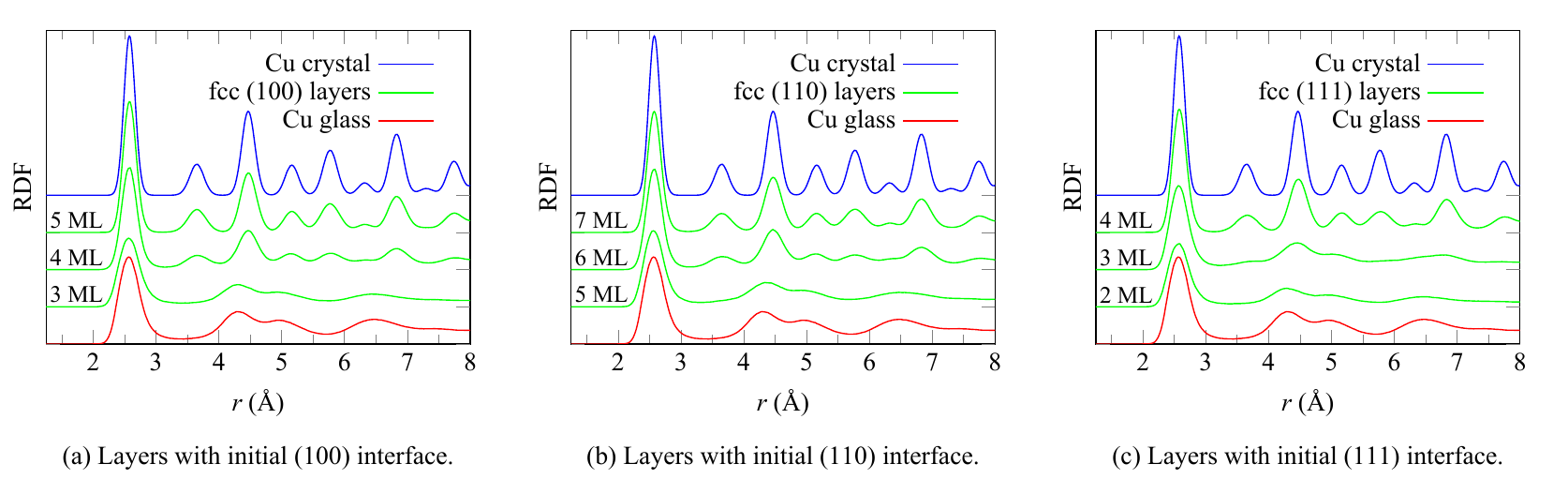}
  \caption{Radial distribution functions of the copper nanolayers in
    several composite systems compared with the reference systems.
    Results of the simulation carried out using the Mendelev
    potential.}
  \label{fig:RDF}
\end{figure*}

\subsection{Sample preparation}

\cuzr{} metallic glasses were prepared by quenching from the melt at
$2000\,\mathrm{K}$ to $300\,\mathrm{K}$ with a cooling rate of
$0.01\,\mathrm{K/ps}$.  This procedure yields a MG with a local
topology matching the experiment \cite{Ritter2011a,Cheng2008}.  Two
glasses with 63108 atoms were prepared, one using the Mendelev
potential and one using the Ward potential.  The final size of the
simulation box is approximately $10 \times 10 \times 10 \,
\mathrm{nm}^3$.  All following steps were carried out twice, once with
the Mendelev and once with the Ward potential.

Copper nanolayers with the appropriate lattice constants at
$300\,\mathrm{K}$ were created with (100), (110), and (111) surfaces.
For each of these, the thickness varies between 2~ML and 15~ML.  To
avoid stresses in the nanolayers after insertion into the glass
matrix, we scaled the $x$ and $y$ dimensions of the glass to fit the
nanolayers exactly and relaxed it again with a barostat applied in $z$
direction at ambient pressure for $1\,\mathrm{ns}$.

The glass was cut at an arbitrary $xy$ plane and the copper layers
were inserted, so that the minimum initial distance between any
nanolayer atom and any matrix atom was at least $1.5\,\mathrm{\AA}$.
The possible consequences of low atomic distance and the resulting
high potential energy at the interface are discussed later in
section~\ref{sec:model:Mendelev}.  To develop a stable interface, the
systems were equilibrated for $1\,\mathrm{ns}$ at $300\,\mathrm{K}$,
again with a barostat applied only in the\linebreak $z$ direction.  We
kept the lateral dimensions constant because any change in them would
be dominated by the relaxation of the MG.  This would induce unwanted
stresses in the nanolayer.  At the end of this procedure, the systems
were completely equilibrated (cf.\
\onlinecite[section~I]{Supplemental}).

\begin{figure*}
  \centering
  \includegraphics{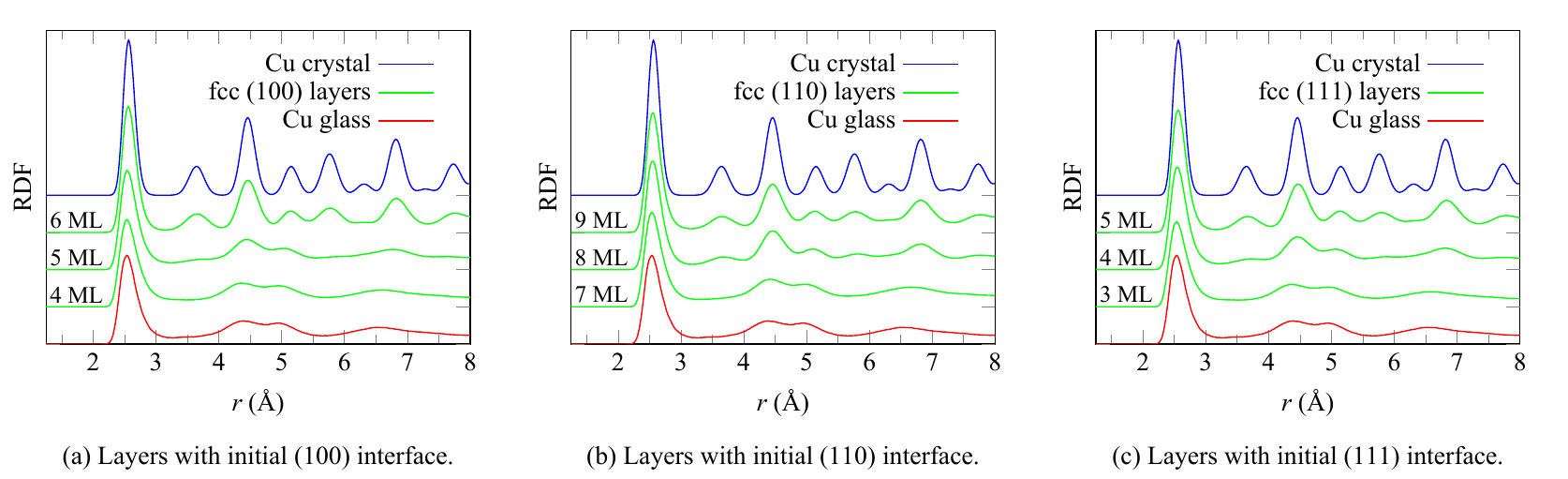}
  \caption{Radial distribution functions of the copper nanolayers in
    several sandwich systems compared with the reference systems.
    Results of the simulation carried out using the Ward potential.}
  \label{fig:RDF-Ward}
\end{figure*}

Further, we created reference systems of crystalline and amorphous
copper phases.  For the crystalline phase, we simply equilibrated fcc
copper at $300\,\mathrm{K}$ to obtain the correct lattice constant.
Bulk amorphous copper was obtained by quenching from the melt at
$2000\,\mathrm{K}$ with very high cooling rates.  We had to employ a
cooling rate of $1\,\mathrm{K/ps}$ for the Mendelev potential and
$25\,\mathrm{K/ps}$ for the Ward potential.  These cooling rates are
approximately the minimum cooling rates needed to avoid
crystallization.  The difference is a result of the different
glass-forming ability of elemental copper in the two potentials.

\subsection{Analysis}

We applied a common neighbor analysis (CNA)
\cite{Honeycutt1987,Stukowski2012} as implemented in \textsc{ovito}
\cite{Stukowski2010} to identify the structure of the nanolayers in
the composite.  The CNA calculates the coordination of all atoms by
examining their neighborhood.  To confirm these results, we calculated
a radial distribution function (RDF) by averaging RDFs calculated for
50 snapshots of the equilibrated systems.  The RDFs were calculated
only for the atoms in the nanolayer.  These RDFs were then compared to
reference RDFs of the copper fcc and copper glass systems.  The
short-range order of the amorphous nanolayers was analyzed using the
Vorono\"i tessellation method implemented in
\textsc{ovito}\cite{Stukowski2010}, which divides the simulation cell
into one polyhedron around each atom
\cite{Voronoi1908,Voronoi1908a,Voronoi1909,Brostow1998}. The polyhedra
are characterized by the Vorono\"i index $\langle n_3, n_4, n_5, n_6
\rangle$, where $n_i$ denotes the number of $i$-edged faces of the
polyhedron.

\section{Results}
\label{sec:results}

\subsection{Mendelev potential}

Figure~\ref{fig:seq-sandwich} shows the time evolution of composite
systems with different nanolayer thickness and the results of the CNA.
Atoms depicted in yellow are nanolayer atoms that are fcc coordinated.
On insertion, the nanolayer had a (100) surface orientation.  We can
see that the nanolayer with 3~ML of copper amorphizes after a short
simulation time, while the copper with 5~ML thickness stays
crystalline. At a thickness of 4~ML a mixed state occurs.  The systems
with different initial orientations show similar behavior.

For obtaining an independent confirmation of the appearance of an
amorphous and crystalline copper phase at different nanolayer
thicknesses, we calculated the RDFs of the nanolayers without
including the glass matrix.  The results are shown in
Figure~\ref{fig:RDF} and compared with the RDFs of the bulk
crystalline and amorphous copper reference phases.  Only three layers
are shown for every initial surface orientation: the thickest
amorphous layer, the thin\-nest crystalline layer, and the layer with a
mixed state.  The results match the CNA.  Amorphous layers show a
similar RDF to the bulk amorphous copper phase, including the
characteristic double peak between $4\,\mathrm{\AA}$ and
$5\,\mathrm{\AA}$.  The RDFs of completely crystalline layers match
the bulk fcc copper, although the signal at large $r$ gets smaller due
to the finite size of the nanolayers.  A mixed state is indicated by
appearance of the second crystalline peak with reduced intensity.
Depending on the fraction of the amorphous phase, the glass
double peak starts separating into the two clearly distinct
crystalline peaks (number three and four).  In the system with initial
(110) surface and a layer thickness of 5~ML, a small trace of
crystalline phase is still detectable as indicated by a slight
increase of the RDF at about $3.5\,\mathrm{\AA}$, the position of the
second crystalline peak.

\subsection{Ward potential}

The simulations using the Ward potential show the same behavior as the
simulations using the Mendelev potential.  The corresponding RDFs of
the copper nanolayers are shown in Figure~\ref{fig:RDF-Ward}.  The
RDFs exhibit the same characteristics, i.e., the gradual disappearance
of the second crystalline peak as well as the appearance of the glass
double peak with decreasing thickness.  The amorphization also occurs
in slightly thicker layers.  For the system with an initial (100)
surface, 4~ML are amorphous, and even the 5~ML system only shows a
small amount of crystallization.  The mixed state for the initial
surfaces (110) and (111) appears at 8~ML and 4~ML, respectively.  The
CNA results agree with the RDFs and are comparable to
Figure~\ref{fig:seq-sandwich}.  They are therefore omitted here.

The simulations with both potentials agree that thin, initially
crystalline nanolayers of copper become amorphous.  Thus, we can
already exclude that amorphization is a result of kinetics in the
deposition process, as the layers were inserted in a crystalline
state.  Still, the creation of the interface may be connected with
local heating, leading to a fast melt-quench process in the smaller
layers.  It is therefore necessary to investigate the thermodynamics
of the system.

\section{Model}
\label{sec:model}

\begin{figure*}
  \centering
  \includegraphics{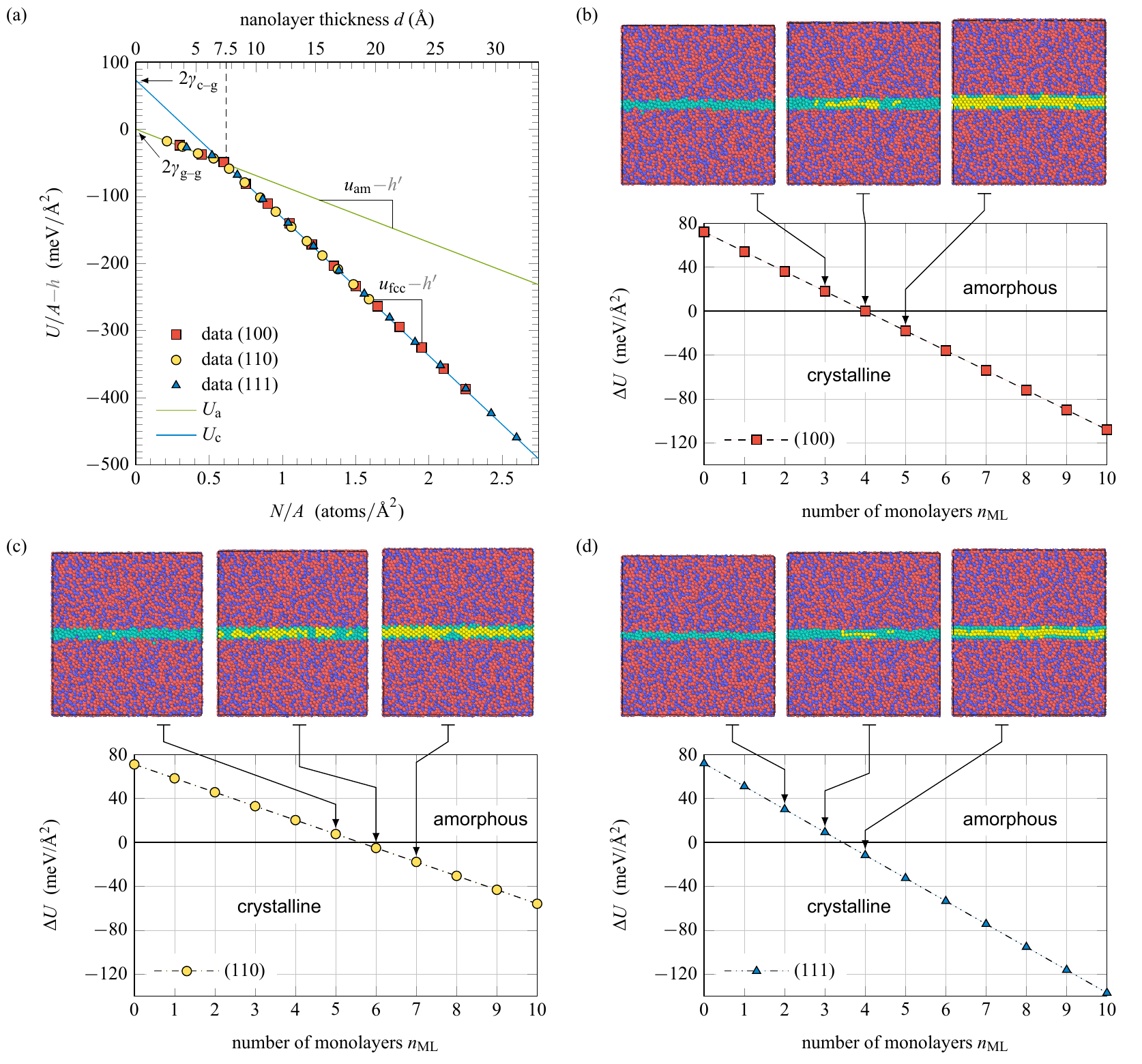}
  \caption{Internal energies of the multilayer systems modelled with the
    Mendelev potential. (a) shows the internal energies extracted from the
    MD simulations as symbols and the linear regression results as
    lines.  For visualization purposes a function $h = h' \cdot N/A$
    with $h' = 3\,\mathrm{eV}$ was subtracted to exaggerate the
    difference in slopes between $U_\mathrm{c}$ and $U_\mathrm{a}$. In
    (b)--(d), $\Delta U$ is plotted as a function of the number of
    monolayers for initial surface orientations (100), (110), and
    (111).  Additionally, snapshots of the MD simulation are added.
    Here, red and blue atoms are copper and zirconium, respectively.
    Green atoms are copper atoms belonging to the nanolayer.  Yellow
    atoms are fcc coordinated.}
  \label{fig:deltaE}
\end{figure*}

An indication for an energetically driven amorphization is already
given by the fact that even initially crystalline nanolayers undergo
a phase transformation to the amorphous state.  To explain the
amorphization of the nanolayers and to test the hypothesis of
energetically driv\-en SSA, we propose a simple thermodynamic model.
We formulate the internal energy $U$ of the composite systems.  In the
given ensemble, the free energy would be the appropriate thermodynamic
potential, but given that the entropy term should favor the
amorphization, we do not artificially increase the driving force for
the amorphization, but rather underestimate it.  The internal energy of a
composite system with an embedded crystalline nanolayer is then
\begin{equation}
  \label{eq:Ec}
  U_\text{c} = \UBMG + N \ufcc + 2 A \gammacg.
\end{equation}
\UBMG{} is the total internal energy of the bulk \cuzr{} glass phase, $N$
is the number of atoms in the nanolayer, \ufcc{} is the internal energy
per atom of the copper fcc crystal.  Additionally, there are two
interfaces, which contribute an energy of $A \gammacg$ each, where $A$
is the interface area.

If the system instead contains an embedded amorphous nanolayer, the
internal energy is expressed as
\begin{equation}
  \label{eq:Ea}
  U_\text{a} = \UBMG + N \uglass + 2 A \gammagg,
\end{equation}
where \uglass{} is the per-atom internal energy of the glassy
nano\-layer and $A \gammagg$ is the glass--glass interface energy.

\begin{figure*}
  \centering
  \includegraphics{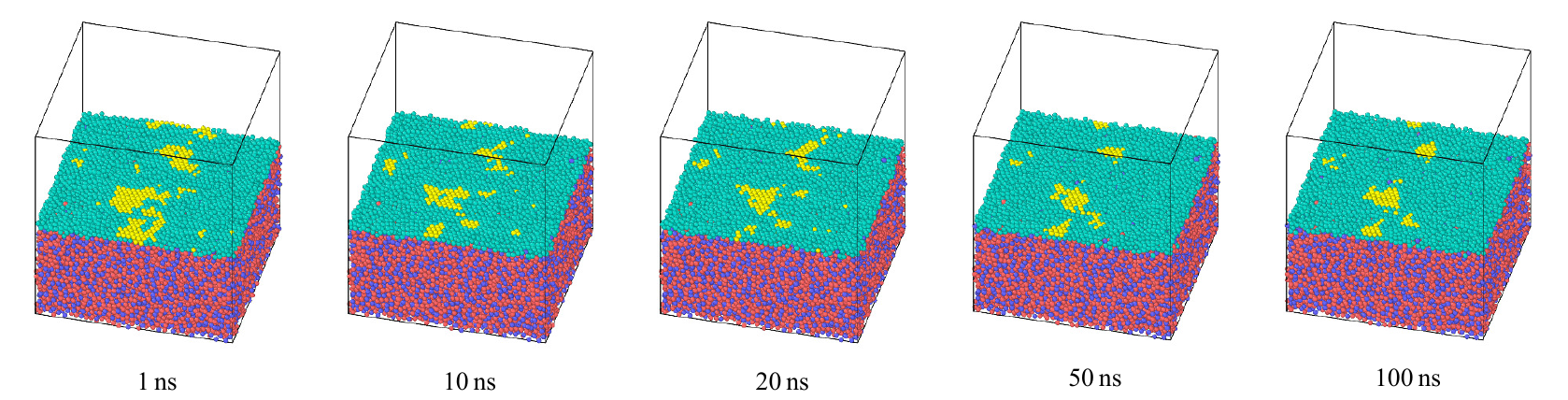}
  \caption{Cut through a nanolayer of 3~ML thickness with an initial
    (111) surface at different time steps.  Even after a simulation
    time of $100\,\mathrm{ns}$, the mixed crystalline/amorphous state
    stays stable.  The simulation was carried out using the Mendelev
    potential.  Copper and zirconium atoms are shown in red and blue,
    respectively.  Copper atoms belonging to the nanolayer are yellow
    if they are fcc coordinated and green otherwise.}
  \label{fig:seq-long}
\end{figure*}
\begin{figure*}
  \centering
  \includegraphics{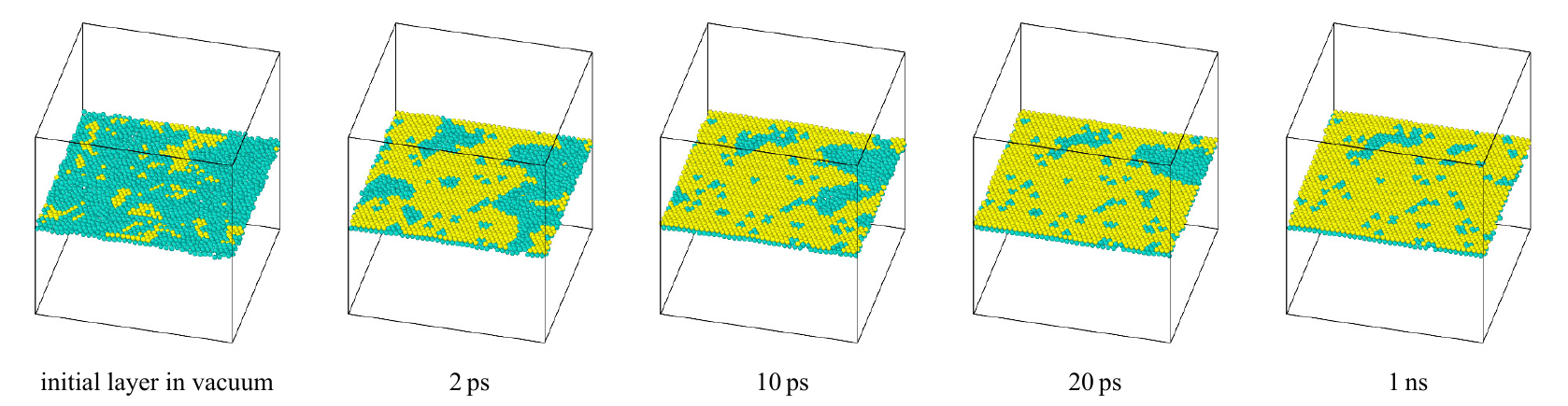}
  \caption{The nanolayer shown in Figure~\ref{fig:seq-long} removed
    from the metallic glass matrix and put in vacuum.  The cut shows
    that the layer crystallizes almost immediately, proving that only
    the glass--glass interface stabilizes the amorphous phase.  Yellow
    atoms are fcc coordinated.}
  \label{fig:in-vacuum}
\end{figure*}

Generally, it is to be expected that the internal energy of a copper
crystal is lower than the internal energy of amorphous copper.  As we
nevertheless see amorphization of pure-metal nanolayers, the reason
must lie in the interface energy.  We propose that the crystal--glass
interface energy \gammacg{} is higher than the glass--glass interface
energy \gammagg{}.  In that case an amorphous nanolayer must be
energetically favorable if its thickness doesn't exceed a critical
value, given that differences in interface energy can compensate the
excess energy of the copper glass phase.  A quantitative measure is
provided by the internal energy difference, here expressed as an
intrinsic quantity independent of the surface area:
\begin{equation}
  \label{eq:deltaE-def}
  \begin{split}
    \Delta U &= (U_\text{c} - U_\text{a}) / A \\
             &= \left( \frac{N}{A} \ufcc + 2\gammacg \right) -
                \left( \frac{N}{A} \uglass + 2\gammagg \right) \\
             &= \frac{N}{A} \Delta u_\text{Cu} + 2 \Delta \gamma.
  \end{split}
\end{equation}
We can see that a negative value of $\Delta U$ signifies a stable
crystalline nanolayer, while a positive value of $\Delta U$ signifies
a stable amorphous nanolayer:
\begin{align*}
  \Delta U(N) &< 0 & &\text{crystalline nanolayer} \\
  \Delta U(N) &> 0 & &\text{amorphous nanolayer}.
\end{align*}
Should the theory hold, we should be able to show that the critical
number of atoms $N_\text{crit}$ at $\Delta U(N_\text{crit}) = 0$ is
the same as observed in the simulation by CNA.  To get a more
descriptive quantity, the number of atoms can easily be converted to
the number of monolayers $n_\text{ML}$ or the thickness of the
nanolayer $d$, as these quantities are proportional:
\begin{equation*}
  \frac{N}{A} \propto n_\text{ML} \propto d.
\end{equation*}

The missing parameters in our model are now $\ufcc$, $\uglass$,
$\gammacg$, and $\gammagg$.  Using equations~\ref{eq:Ec}
and~\ref{eq:Ea}, the internal energies of the different layer phases can
be obtained from the relations
\begin{equation}
  \label{eq:lin-eq-bulk-energy}
  \ufcc   = \frac{\mathrm{d}U_\text{c}}{\mathrm{d}N}
  \quad
  \text{or}
  \quad
  \uglass = \frac{\mathrm{d}U_\text{a}}{\mathrm{d}N},
\end{equation}
respectively.  Alternatively, it would be conceivable to just use the
internal energies of the bulk copper reference systems.  The problem
would be that the amorphous phase in the nanolayer is not necessarily
the same as in the bulk.  The bulk amorphous copper is quenched with
very high cooling rates and has therefore more similarity to the melt.
The amorphous copper phase in the nanolayer may exhibit different
short-range order as it is allowed to relax to a low-energy state.
Furthermore, the ratio of interface to volume is very high, which
means that the nanolayer structure is highly influenced by interface
contributions.  To calculate the interface energy, the internal energy
\UBMG{} is first taken from the pure \cuzr{} glass sample before
embedding the copper nanolayer.  This allows to calculate the
interface energies either by directly using equations~\ref{eq:Ec}
and~\ref{eq:Ea}, or by subtracting \UBMG{} from the $U$-axis intercept
of the $U(N)$ curves.  Both should yield the same result.  We assume here
that the interface energy is constant, see
\onlinecite[section~II]{Supplemental} for proof.

\vspace{-\baselineskip}
\subsection{Mendelev potential}
\label{sec:model:Mendelev}
\begin{figure*}
  \centering
  \includegraphics{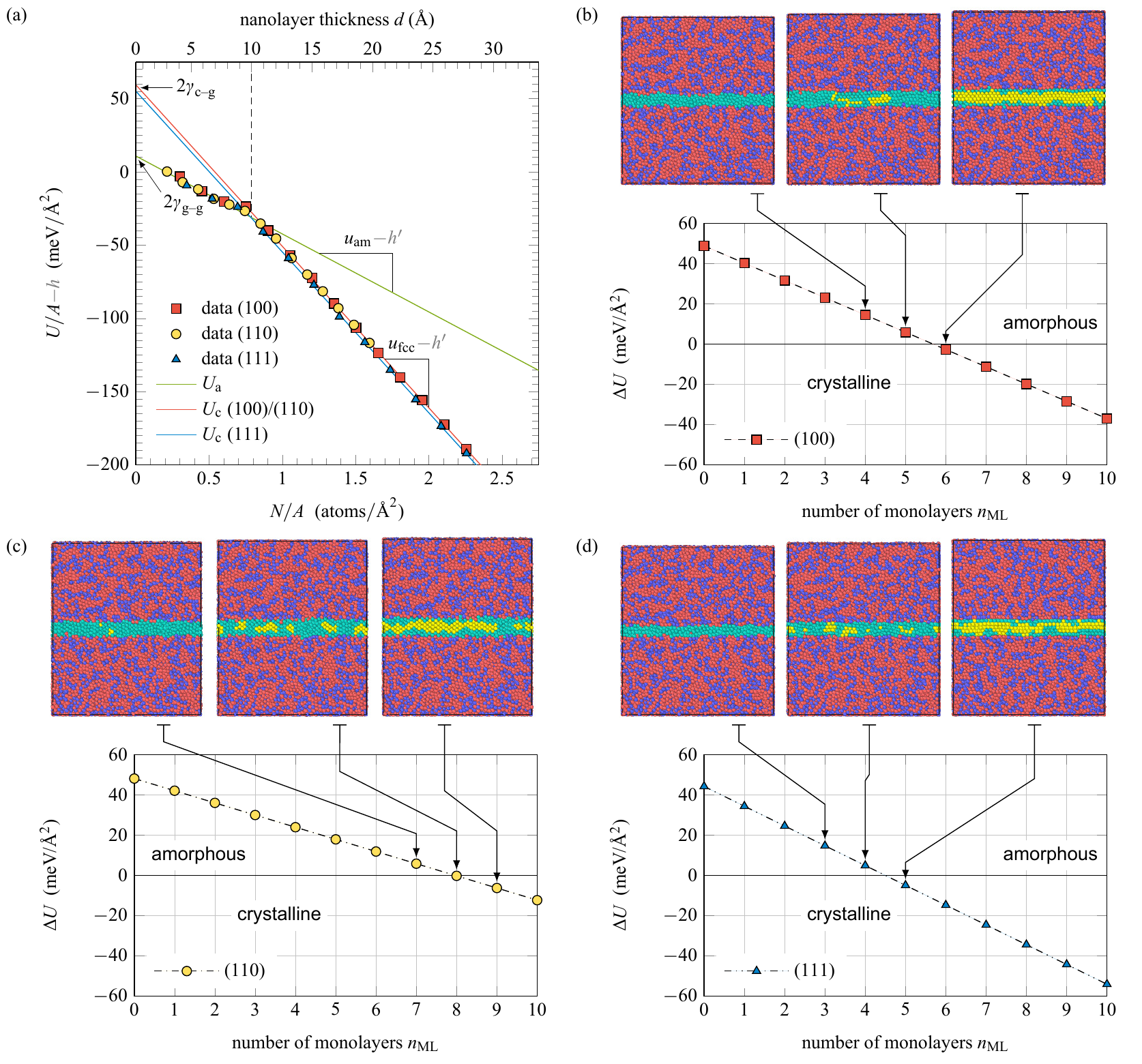}
  \caption{Internal energies of the multilayer systems modelled with the
    Ward potential. (a) shows the internal energies extracted from the
    MD simulations as symbols and the linear regression results as
    lines.  For visualization purposes a function $h = h' \cdot N/A$
    with $h' = 3.35\,\mathrm{eV}$ was subtracted to exaggerate the
    difference in slopes between $U_\mathrm{c}$ and $U_\mathrm{a}$. In
    (b)--(d), $\Delta U$ is plotted as a function of the number of
    monolayers for initial surface orientations (100), (110), and
    (111).  Additionally, snapshots of the MD simulation are added.
    Here, red and blue atoms are copper and zirconium, respectively.
    Green atoms are copper atoms belonging to the nanolayer.  Yellow
    atoms are fcc coordinated.}
  \label{fig:deltaE-Ward}
\end{figure*}

Figure~\ref{fig:deltaE}a shows the internal energies $U_\text{c}$ and
$U_\text{a}$ as a function of the number of atoms.  All values are
normalized to the interface area and \UBMG{} is already subtracted.
The symbols show the internal energies extracted from the MD simulation,
while the lines show the linear regression.  The numerical data is
listed in Table~\ref{tab:data}.  We note that all crystalline
nanolayers have (approximately) the same interface energy in the
Mendelev potential.  The graph shows that the glass--glass interface
energy is lower than the crystal--glass interface (in fact it is zero,
which will be discussed in detail in section~\ref{sec:struc-ener}),
which fits the assumptions of our model.  This lowered $\gammagg$
favors a glassy nanolayer up until approximately $0.6\,\mathrm{atoms/\AA^2}$,
which corresponds to a critical thickness of about
\begin{equation}
  d_\text{crit} \approx 7.5\,\text{\AA}.
\end{equation}
The thickness $d_\text{crit}$ can only be given approximately, due to
the different densities of the two phases and the rough interface.
Figures~\ref{fig:deltaE}b--d show $\Delta U$ as a function of the
number of monolayers.  The direct comparison reveals that the
calculated energy differences and the solid-state amorphization
observed by the CNA method agree very well.  At the critical thickness
$\Delta U=0$ we observe a mixed crystalline/amorphous nanolayer.  The
figures show that the transition is not as sharp as our model assumes,
a partly crystalline layer also exists for $\Delta U$ slightly greater
than zero.  That is a result of omitting a description of the
two-phase region in the model; entropy and the additional interfaces
play a role here.  Nonetheless, the critical thickness is correctly
reproduced without these complications.  A further comparison with the
RDFs in Figure~\ref{fig:RDF} leads to the same conclusions.  While the
good fit of the model with simulation data supports the conclusion
that the SSA is due to energetic reasons, we also investigated the
influence of the simulation setup on the amorphization of the
nanolayers in \onlinecite[section~III]{Supplemental}. We find that the
interface creation leads to a heat spike, but that this only serves as
activation energy for the SSA process. Crystalline layers with a
thickness below $d_\text{crit}$ can be produced, but are not
energetically favorable.

In order to verify that the amorphous phase is indeed stable over a
long time scale and that this is a result of the glass--glass
interface, we conducted two further simulations.  In the first
simulation, we simply took a composite system with a mixed
crystalline/amorphous state in the copper nanolayer and let the
simulation run for $100\,\mathrm{ns}$.  If the amorphous phase is only
produced by, e.g., stress in the initial system after insertion of the
nanolayer, the crystalline phase should start growing again over the
longer time frame.  The simulation results are depicted in
Figure~\ref{fig:seq-long} and show that the mixed state is stable, as
predicted by our model.

A direct proof that the solid-state amorphization is due to the
presence of a glass--glass interface was obtained by removing the glass
matrix.  The results are shown in Figure~\ref{fig:in-vacuum}.  The
free amorphous layer crystallizes almost immediately, as would be
expected.

\subsection{Ward potential}
\begin{figure}[b]
  \centering
  \includegraphics[]{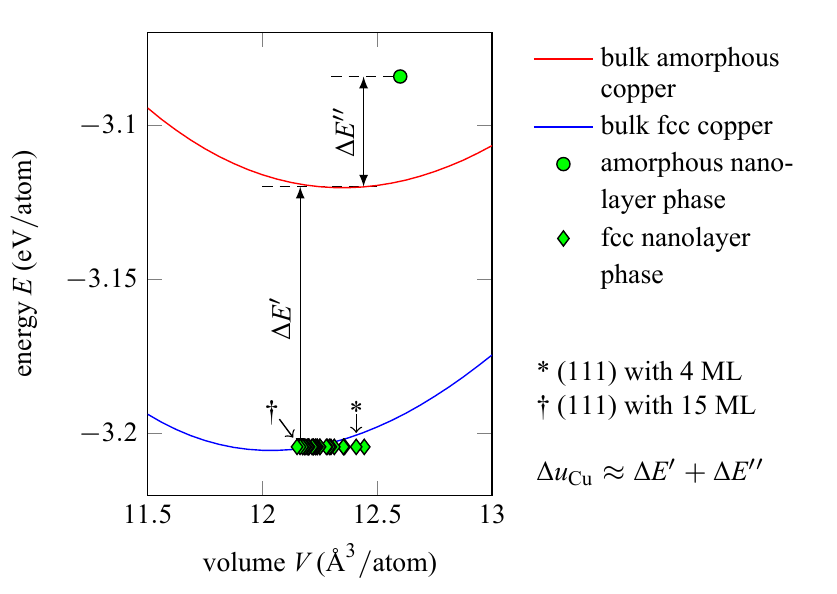}
  \caption{Energy--volume curves for amorphous and crystalline copper
    and data points for amorphous and crystalline nanolayers.  Atomic
    volumes of the nanolayers were obtained by a Gauss fit to the
    Vorono\"i volume distribution.  The crystalline nanolayers vary in
    density with their thickness, approaching the equilibrium volume
    of an fcc crystal at higher thickness.}
  \label{fig:EV-Cu}
\end{figure}

Figure~\ref{fig:deltaE-Ward}a shows the internal energies as
calculated with the Ward potential as a function of the number of
atoms.  All values are again normalized to the interface area and
\UBMG{} is already subtracted.  The symbols show the internal energies
extracted from the MD simulation, while the lines show the linear
regression.  In the Ward potential the (111) interface has a slightly
lower interface energy than the (100) and (110) interfaces, which are
approximately the same (see also Table~\ref{tab:data}).  As with the
Mendelev potential, the glass--glass interface energy is lower than
the crystal--glass interface energy, favoring an amorphous nanolayer
up to the critical thicknesses
\begin{align}
  d_\text{crit}^{(100)} &\approx 10.8\,\text{\AA}, \\
  d_\text{crit}^{(110)} &\approx 10.7\,\text{\AA}, \text{ and} \\
  d_\text{crit}^{(111)} &\approx \phantom{0}9.9\,\text{\AA}.
\end{align}
The difference in critical thickness is a result of different
$\gammacg$ for the three surface orientations.  The transition
thickness is higher than in the Mendelev potential despite a smaller
$\Delta \gamma$, as the excess energy of the amorphous phase is lower.
By plotting $\Delta U$ as a function of the number of monolayers, a
direct comparison to CNA and RDF results is possible.
Figures~\ref{fig:deltaE-Ward}b--d show $\Delta U(n_\text{ML})$
compared with snapshots from the simulation.  Again, a good match
between the nanolayer phases shown in the snapshots and the predicted
critical thickness is visible.  For the same reasons as stated
earlier, a mixed state occurs.

All in all, the results using both potentials agree qualitatively and
support our thermodynamic model.  Therefore, a purely kinetic reason
for the amorphous nanolayers can be ruled out and an energetic picture of
solid-state amorphization can be supported.

\begin{figure}[t]
  \centering
  \includegraphics[]{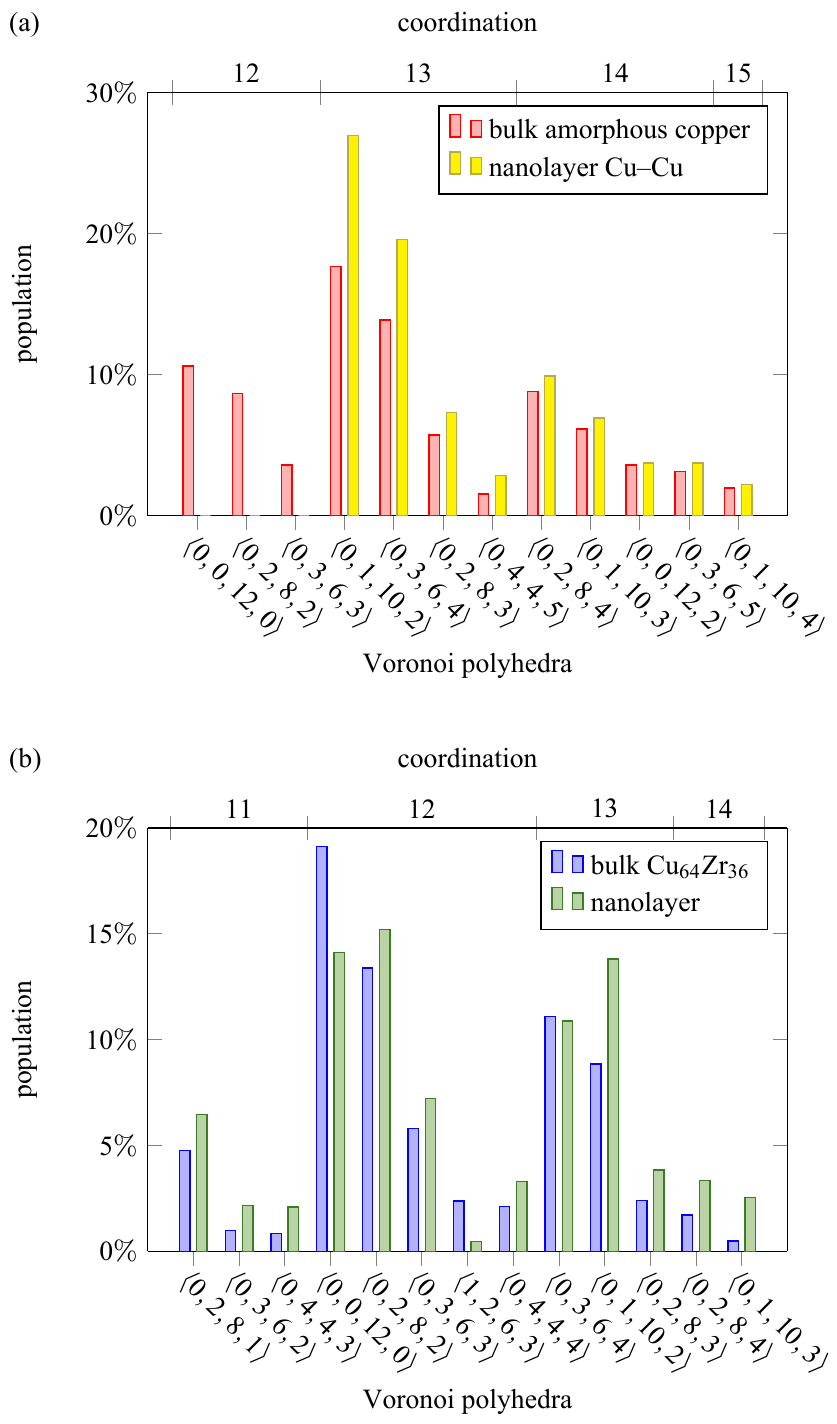}
  \caption{Vorono\"i analysis of the amorphous nanolayers in systems
    simulated using the Mendelev potential.  In (a) the bulk amorphous
    copper phase is compared with the amorphous copper phase in the
    nanolayers.  Only those nanolayer atoms were included, which were
    surrounded solely by other copper atoms.  In (b) all nanolayer
    atoms (including those with zirconium neighbors) were considered
    and compared to \cuzr{} bulk.}
  \label{fig:voronoi}
\end{figure}

\section{Structure and Energy}
\label{sec:struc-ener}

\subsection{Mendelev}

We compared the energy of the amorphous nanolayer copper phase with
the reference bulk amorphous copper phase.  The energy--volume curve
in Figure~\ref{fig:EV-Cu} shows that the nanolayer phase is
energetically higher than the bulk phase.  Additionally, an expansion
of the crystalline layer at low thicknesses is visible, possibly due
to interface stress.  Both the peculiarity of the zero glass--glass
interface energy, as well as the higher excess energy of the amorphous
phase can be linked to the structure of the phase.
Figure~\ref{fig:voronoi}a shows the Vorono\"i statistics of those copper
atoms in the nanolayer, that are only surrounded by other copper
atoms.  This allows a comparison to bulk amorphous copper: in contrast
to the bulk phase, the nanolayer phase contains no twelve-fold
coordinated atoms.  This higher energy structure is stabilized by the
interface, as shown in Figure~\ref{fig:voronoi}b: the Vorono\"i
statistics of the whole nanolayer (including copper atoms that have
zirconium neighbors) are very similar to the bulk \cuzr.  This reduces
the interface energy to almost zero.
\pagebreak
\subsection{Ward}
\begin{figure}[t]
  \centering
  \includegraphics[]{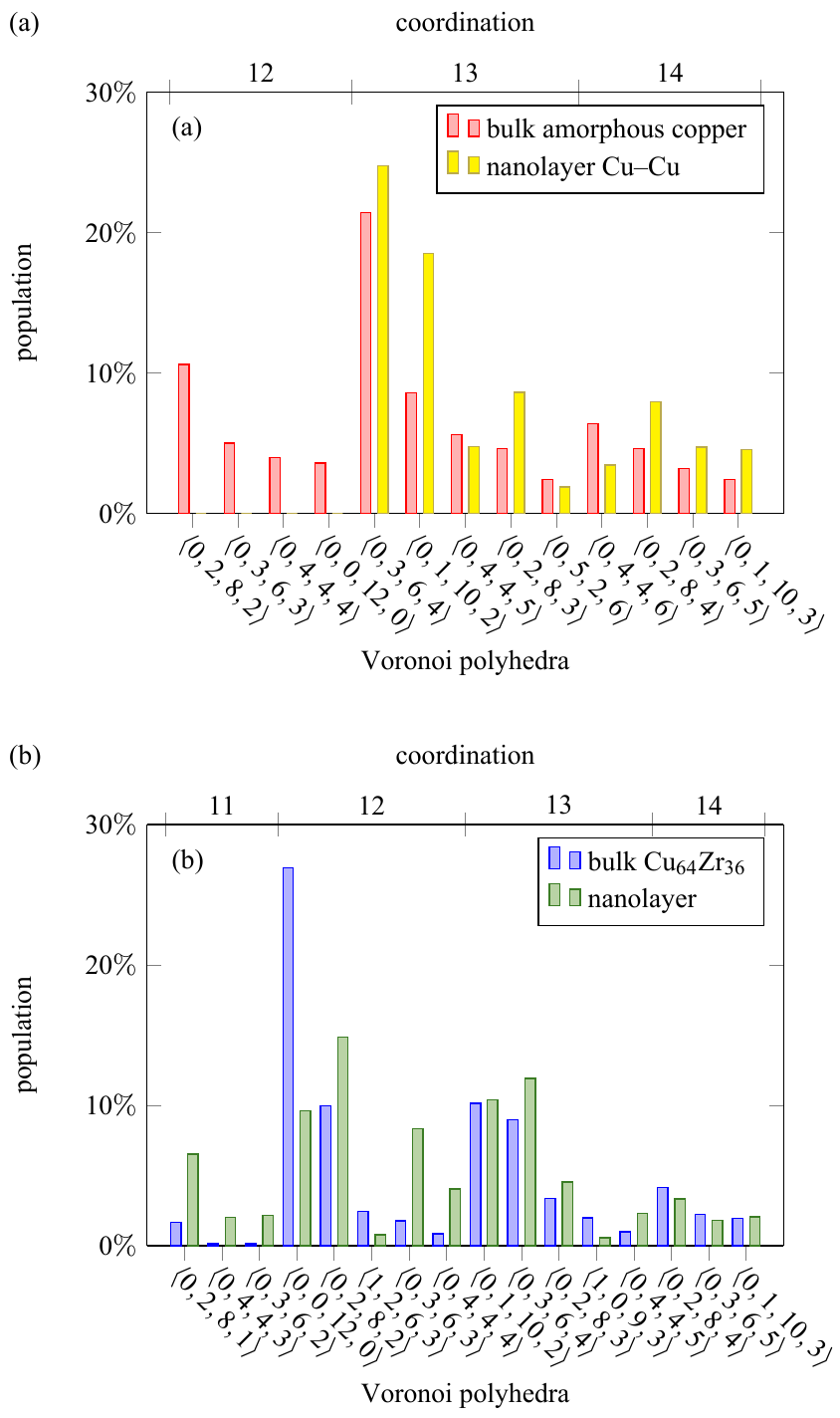}
  \caption{Vorono\"i analysis of the amorphous nanolayers in systems
    simulated using the Ward potential.  In (a) the bulk amorphous
    copper phase is compared with the amorphous copper phase in the
    nanolayers.  Only those nanolayer atoms were included, which were
    surrounded solely by other copper atoms.  In (b) all nanolayer
    atoms (including those with zirconium neighbors) were considered
    and compared to \cuzr{} bulk.}
  \label{fig:voronoi-Ward}
\end{figure}

In the Ward potential the amorphous nanolayer also features a
structure with different energy than the bulk amorphous copper
($-3.39\,\mathrm{eV/atom}$ for the reference system,
$-3.40\,\mathrm{eV/atom}$ for the amorphous nanolayer).  The
explanation can again be found in the Vorono\"i statistics of nanolayer
copper atoms surrounded completely by other copper atoms:
Figure~\ref{fig:voronoi-Ward}a shows that the twelve-fold coordinated
atoms are missing again.  The comparison of the Vorono\"i statistics of
the whole nanolayer, though, show a difference to the \cuzr{} MG,
especially concerning the $\langle 0,0,12,0 \rangle$, $\langle 0,2,8,2
\rangle$, and $\langle 0,3,6,3 \rangle$ polyhedra
(Figure~\ref{fig:voronoi-Ward}b).  This leads to a small but non-zero
glass--glass interface energy.

\begin{table*}
  \centering
  \caption{Internal energies and interface energies extracted from MD
    simulations and predicted critical thicknesses.}
  \label{tab:data}
  \begin{ruledtabular}
  \newcolumntype{d}[1]{D{.}{.}{#1}}
  \begin{tabular}{ccd{2.2}d{2.2}d{2.2}d{2.1}d{2.1}d{2.1}d{1.3}d{2.1}}
    \multirow{2}{*}{Potential}
    & \multirow{2}{1.5cm}{\centering Initial interface}
    & \multicolumn{1}{c}{\ufcc}
    & \multicolumn{1}{c}{\uglass}
    & \multicolumn{1}{c}{$\Delta u_\text{Cu}$}
    & \multicolumn{1}{c}{\gammacg}
    & \multicolumn{1}{c}{\gammagg}
    & \multicolumn{1}{c}{$\Delta \gamma$}
    & \multicolumn{1}{c}{$N_\text{crit}/A$}
    & \multicolumn{1}{c}{$d_\text{crit}$}
    \\
    \cline{3-5} \cline{6-8}
    &
    & \multicolumn{3}{c}{($\mathrm{eV}/\text{atom}$)}
    & \multicolumn{3}{c}{($\mathrm{meV}/\text{\AA}^2$)\rule{0pt}{10.5pt}}
    & \multicolumn{1}{c}{($\text{atoms}/\text{\AA}^2$)}
    & \multicolumn{1}{c}{(\AA)} \\
    \colrule
    \multirow{3}{*}{Mendelev}
      & (100)
      & -3.20 & -3.08& -0.12
      & 36.5  & 0.4  & 36.1
      & 0.601
      & 7.5
    \\
      & (110)
      & -3.20 & -3.08 & -0.12
      & 35.5  & 0.1   & 35.4
      & 0.592
      & 7.5
    \\
      & (111)
      & -3.20 & -3.08 & -0.12
      & 36.0  & -0.5  & 36.5
      & 0.606
      & 7.5
    \\
    \colrule
    \multirow{3}{*}{ Ward}
      & (100)
      & -3.46 & -3.40 & -0.06
      & 29.7  &  5.4  & 24.4
      & 0.855
      & 10.8
    \\
      & (110)
      & -3.46 & -3.40 & -0.06
      & 29.5  &  5.5  & 24.0
      & 0.846
      & 10.7
    \\
      & (111)
      & -3.46 & -3.40 & -0.06
      & 27.6  &  5.5  & 22.1
      & 0.781
      & 9.9
    \\
  \end{tabular}
  \end{ruledtabular}
\end{table*}

\section{Conclusions}
\label{sec:conclusions}

Using MD simulations, we observed the amorphization of elemental
copper nanolayers embedded in a \cuzr{} metallic glass if the layer
thickness stays below a critical value. This is in accordance with
experimental results, which report thin amorphous iron nanolayers
embedded in Co$_{75}$Fe$_{12}$B$_{13}$ \cite{Ghafari2012}.  We could
show that the amorphization is not a kinetic effect due to deposition,
as our simulations start from a crystalline state.  Rather, the
glass--glass interface energy is significantly lower than the
crystal--glass interface energy, which stabilizes the amorphous copper
phase.  This solid-state amorphization is similar to the case at
heterogeneous crystal interfaces, except that in our case the reduced
glass--glass interface energy is sufficient to induce amorphization.
At a critical layer thickness, which is on the order of a nanometer, a
mixed crystalline/amorphous state appears.  This state is also stable
over longer times, which further supports the picture of solid-state
amorphization: if the amorphous state is only a result of stresses in
the initial setup, the crystallites in the layer would grow again with
time.  They instead keep their size.
Analysis of the amorphous structure in the nanolayer further confirms
that the interface energy is a dominating factor in the structure of
thin nanolayers: if it can be reduced, the amorphous layer can even be
driven into a state with a higher bulk energy than a quenched melt.
Technological applications for glass--glass composite systems have
already been proposed in the realm of magnetic tunnel junctions
\cite{Gao2009,Ghafari2012} and could benefit from further research
into different multilayer systems.

\begin{acknowledgements}
  We would like to thank Mohammad Ghafari for many helpful
  discussions.  The authors gratefully acknowledge financial support
  by the Deut\-sche For\-schungs\-ge\-mein\-schaft (DFG) through
  project grant no.\ AL 578/13-1, as well as a DAAD-PPP travel grant to
  Finland.  We also acknowledge the computing time granted by the John
  von Neumann Institute for Computing (NIC) and provided on the
  supercomputer JUROPA at J\"ulich Supercomputing Centre (JSC).
  Computation time was also made available by the State of Hesse on
  the Lichtenberg Cluster at TU Darmstadt.
\end{acknowledgements}

\bibliographystyle{apsrev4-1}
\bibliography{literatur.bib}

\balancecolsandclearpage


\section*{Supplementary material}

# Supplemental Material on the Simulation Setup
## Solid-state amorphization of Cu nanolayers embedded in a $Cu_{64}Zr_{36}$ glass

This supplemental material discusses some additional technical aspects of the simulation setup in detail. In the first section, we show that the systems are equilibrated. We demonstrate in the second section that the interface energy is a constant independent of the layer thickness. In the final section, we discuss the kinetics during interface creation and show that they do not influence the thermodynamics of the system.

## I. EQUILIBRATION OF GLASS MATRIX AND COMPOSITES

To extract the correct energies from the molecular dynamics (MD) simulations, it is necessary to make sure that the systems are equilibrated, i.e., that the energy of the system is constant in time. The composites were equilibrated for 1 ns at 300 K, which is sufficient to equilibrate the composite systems. This is shown exemplarily for the Mendelev potential, for systems with initial (100)

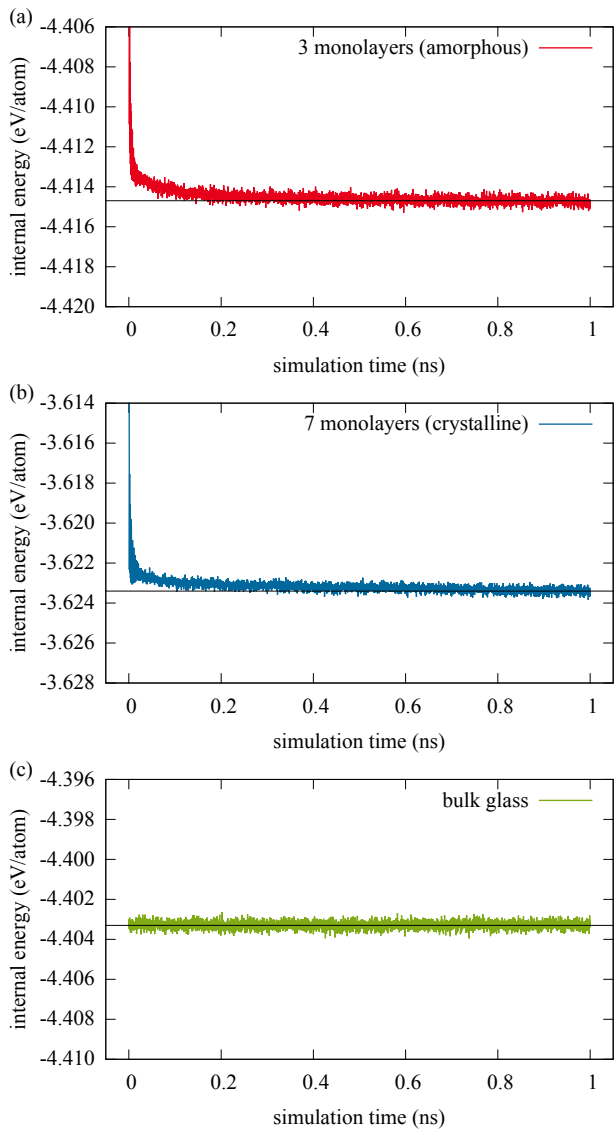

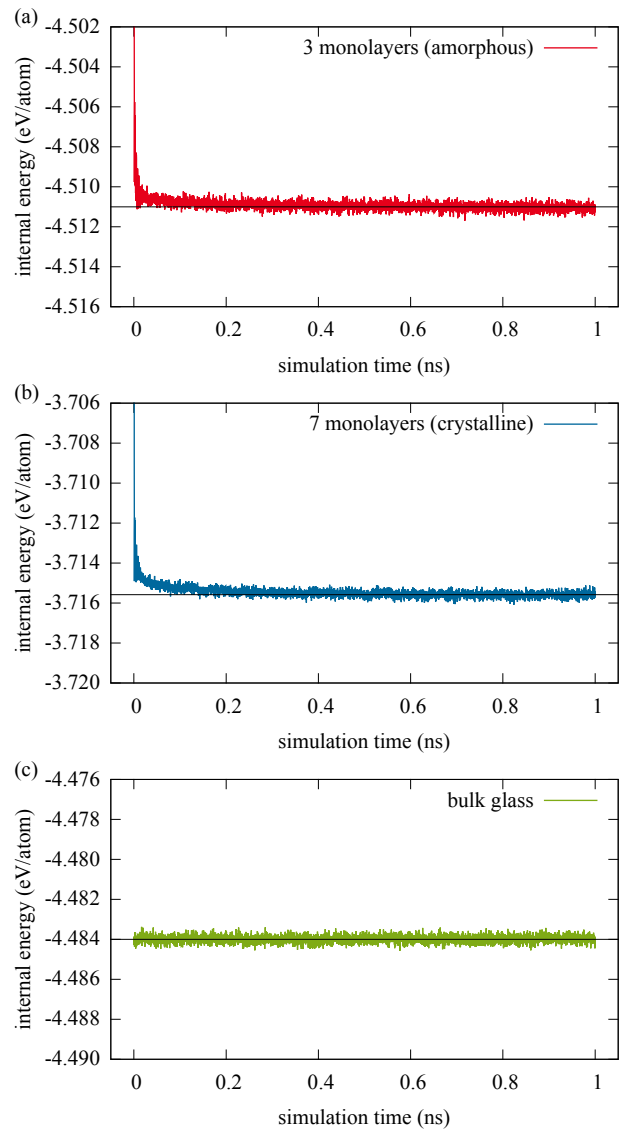

FIG. 1. Internal energy over time for systems simulated with the Mendelev potential. (a–b) Internal energy of two metallic glass–copper composite systems. (c) Internal energy of the pure metallic glass.

FIG. 2. Internal energy over time for systems simulated with the Ward potential. (a–b) Internal energy of two metallic glass–copper composite systems. (c) Internal energy of the pure metallic glass.

interface and 2 monolayers of copper (Figure 1a), as well as 15 monolayers of copper (Figure 1b). For the Ward potential, the same is shown in Figures 2a and 2b. To ensure that the energy of the glass matrix stays constant over the simulation time, we ran simulations with only the glass matrix for 1 ns. These are shown in Figures 1c and 2c and show that the glass is already equilibrated and does not change in internal energy.

## II. INDEPENDENCE OF THE INTERFACE ENERGY FROM LAYER THICKNESS

Our model assumes an interface energy independent of the layer thickness. To prove that this assumption is correct, we additionally performed simulations on a system with 100 monolayers of copper using the Mendelev and the Ward potential. We inserted the results of this simulation into plots of the internal energy as a function of the number of layer atoms, both normalized to the interface area. Figures 3 and 4 show that the composites with very big copper layers have the same energy $U_c$ as predicted by extrapolation using our model (line). Furthermore, we calculated $u_{fcc}$ for the copper nanolayer atoms far from the interface, as well as the per-atom energy of the metallic glass matrix far from the interface and compared them to the values from Table I in the main paper. Tables I and II list the results and clearly show that the results in a system with "bulk" copper agree with the results for composites with thin copper layers. The small deviation of the interface energy, as well as the deviation of $U_{MG}$ for the Ward potential, lie inside the range of numerical errors. The independence of the interface energy from the layer thickness is already a reasonable assumption, given the linear behavior of $U_c$ and $U_a$. With these additional data, the assumption is proven.

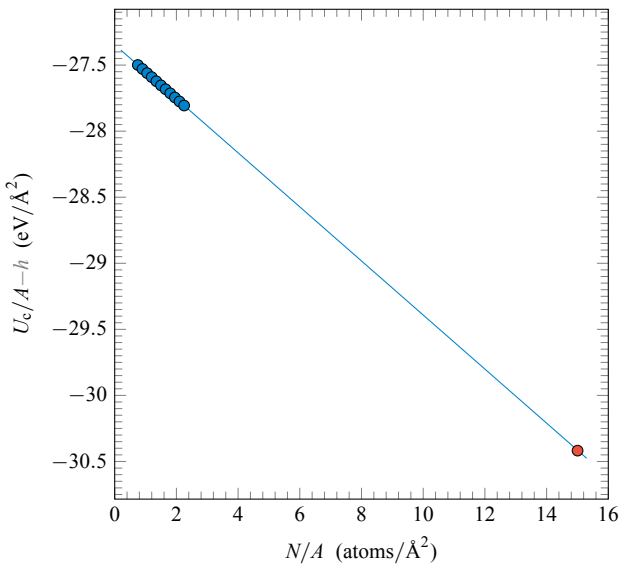

FIG. 3. Internal energy over number of atoms for crystalline nanolayers with initial (100) surface in the Mendelev potential. The blue circles show the results from Figure 4a in the main paper. The red circle is the system with 100 monolayers of copper. To exaggerate the slope for visualization purposes, we subtracted a function $h = 3\,\mathrm{eV} \cdot N/A$ from $U/A$.

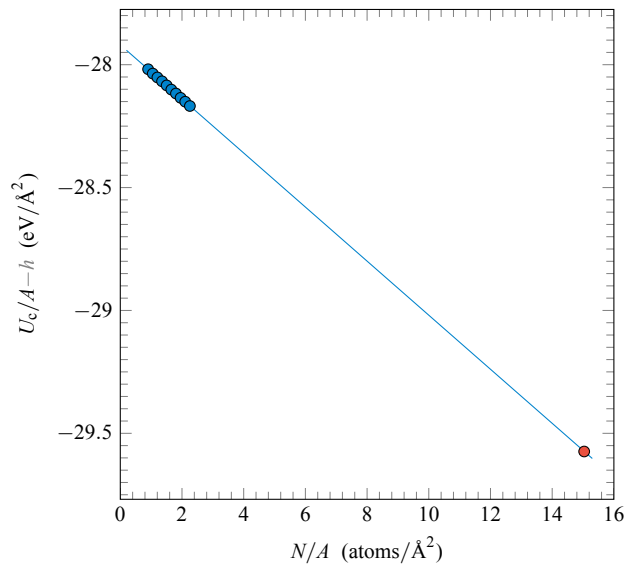

FIG. 4. Internal energy over number of atoms for crystalline nanolayers with initial (100) surface in the Ward potential. The blue circles show the results from Figure 7a in the main paper. The red circle is the system with 100 monolayers of copper. To exaggerate the slope for visualization purposes, we subtracted a function $h = 3.35\,\mathrm{eV} \cdot N/A$ from $U/A$.

TABLE I. Comparison of the interface energy in a composite system with initial (100) surface modelled using the Mendelev potential.

|  | $u_{fcc}$ (eV/atom) | $U_{MG}/N_{MG}$ (eV/atom) | $\gamma_{c-g}$ (meV/Å²) |
|---|---|---|---|
| 5–15 layers | −3.20 | −4.40 | 36.5 |
| 100 layers | −3.20 | −4.40 | 35.6 |

TABLE II. Comparison of the interface energy in a composite system with initial (100) surface modelled using the Ward potential.

|  | $u_{fcc}$ (eV/atom) | $U_{MG}/N_{MG}$ (eV/atom) | $\gamma_{c-g}$ (meV/Å²) |
|---|---|---|---|
| 6–15 layers | −3.46 | −4.48 | 29.7 |
| 100 layers | −3.46 | −4.49 | 29.3 |

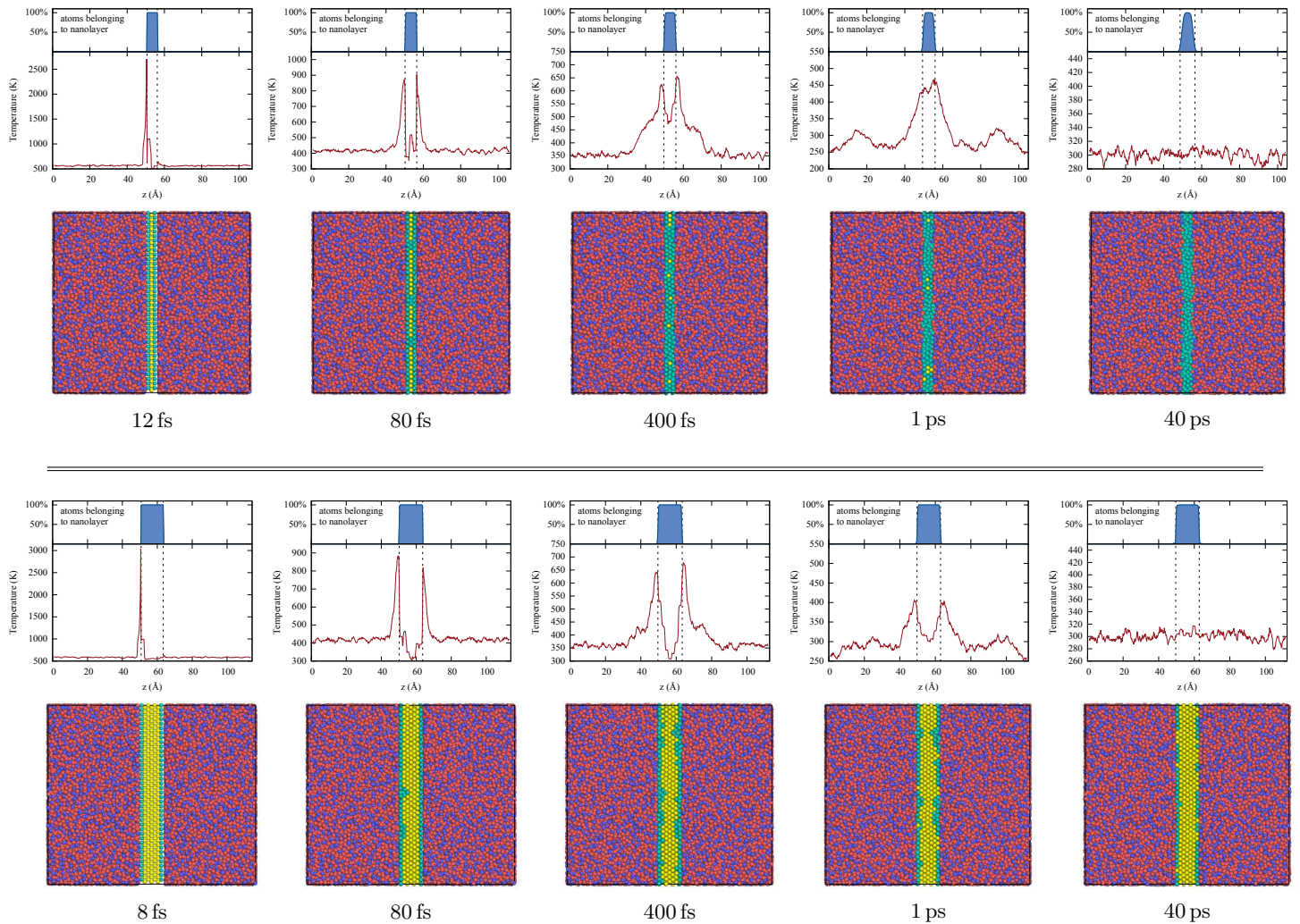

FIG. 5. Temperature profiles for composite systems simulated using the Mendelev potential. The top row shows a system with 3 monolayers, the bottom row a system with 7 monolayers of copper. The graphs are produced by averaging over 2 Å thick slices, of which there are 500 in $z$ direction. The top part of the graphs show the ratio of layer atoms to matrix atoms in the current slice to indicate the position of the two phases and their interface. The lower part of the graphs show the temperature in the current slice. Below the graphs, the corresponding snapshots of the composite systems are shown. The leftmost snapshots are always from the simulation step with the highest temperature spike. *Note:* After creating the composite systems, we initialize the momenta from a Maxwell-Boltzmann distribution. For the first few steps the distribution between kinetic and potential energy is not equilibrated, which leads to an instantaneous temperature higher than 300 K in the glass matrix.

## III. CREATION OF THE COMPOSITE SYSTEMS AND KINETICS AT THE INTERFACE

After inserting the copper nanolayers, some atoms from the nanolayer may have a small distance to atoms from the metallic glass matrix. This could lead to a high potential energy and consequently initial temperature spikes at the interfaces, which may be responsible for the amorphization of the nanolayers. In fact, for some time at the beginning of the simulation, temperature spikes appear mainly on the matrix side of the interface (Figures 5 and 6). In case of the systems with nanolayers thick enough to stay crystalline, the temper-

ature inside of the layer stays comparatively low, while there is an increase of temperature in the layers when they become amorphous. While this may be a sign of the amorphous phase being the result of a fast melt-quench process, it may also be that the initial heat spike only provides an activation energy for an energetically favorable solid state amorphization. To test this, we used a modified simulation setup: The system was prepared in the same way as described in the main paper, but the nanolayers were held fixed in the crystalline state for 1 ns. This avoids penetration of heat into the layer. After that, the nanolayers were released and the whole system equilibrated normally for an additional 1 ns. Additionally, we

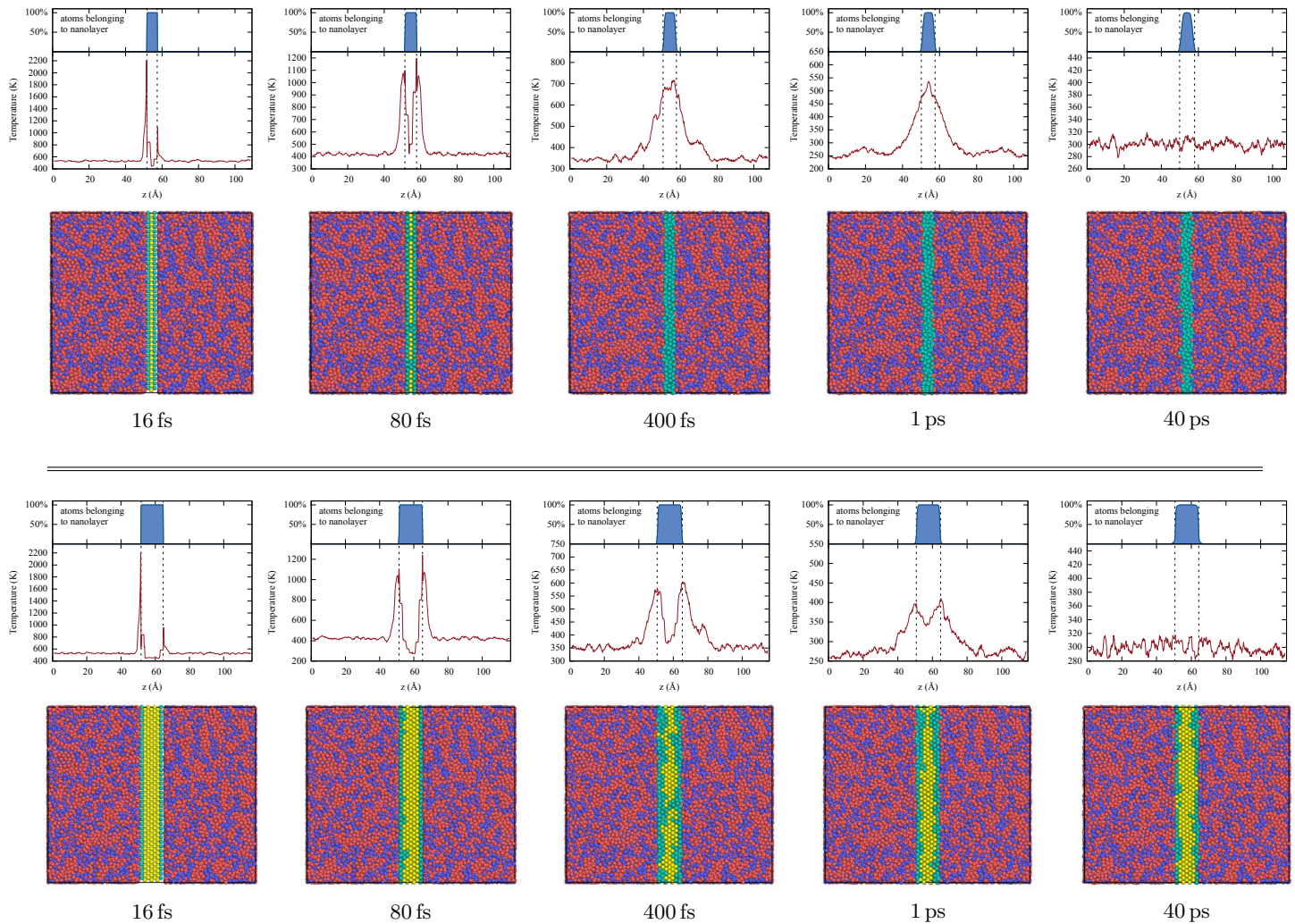

FIG. 6. Temperature profiles for composite systems simulated using the Ward potential. The top row shows a system with 3 monolayers, the bottom row a system with 7 monolayers of copper. The graphs are produced by averaging over 2 Å thick slices, of which there are 500 in $z$ direction. The top part of the graphs show the ratio of layer atoms to matrix atoms in the current slice to indicate the position of the two phases and their interface. The lower part of the graphs show the temperature in the current slice. Below the graphs, the corresponding snapshots of the composite systems are shown. The leftmost snapshots are always from the simulation step with the highest temperature spike.

ran molecular statics simulations on the initial system, which also avoids heating the system locally. The results are shown in Figures 7 and 8. The molecular statics simulations never lead to an amorphization of the copper layer. The reason for that is simply that a phase transformation from fcc copper to amorphous copper needs a certain activation energy. We can also see that the energy of the statically relaxed systems is always much higher than the MD-equilibrated systems. This shows that even the creation of a low-energy crystal–glass interface needs some activation energy. When holding the copper layers fixed, very small layers will still become amorphous, while the amorphization is suppressed in thicker layers. These crystalline nanolayers are metastable, as can be seen in the graphs: Their energy agrees with the prediction of

our model (blue line), which also states that $\Delta U > 0$ in this case. Only above the critical thickness, at the crossover of $U_a$ (dashed line) and $U_c$ (blue line), are the crystalline nanolayers stable again. This is in complete agreement with our model and shows that initially fixing the copper layers only kinetically inhibits the phase transformation on the simulated time scales.

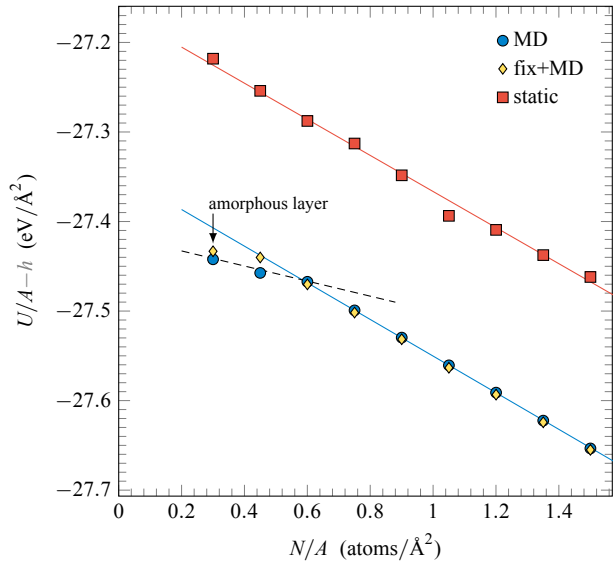

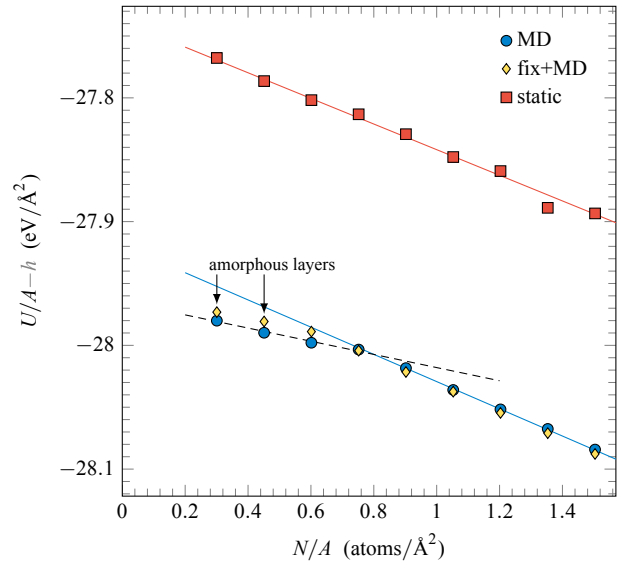

FIG. 7. Internal energy of the composite systems as a function of the number of atoms in the nanolayer. This graph shows nanolayers with initial (100) surface in the Mendelev potential. Data was normalized to the interface area and a function $h = 3\,\mathrm{eV} \cdot N/A$ was subtracted from $U/A$ to exaggerate the difference in slopes for visualization purposes. The blue circles show the results from Figure 4a in the main paper. The yellow diamonds show the results of an MD simulation where the layer atoms are first held in position, followed by a normal MD simulation. Except where marked, the nanolayers stayed crystalline in these simulations. Red squares show a simple molecular statics calculation plus a $2 \times \frac{3}{2}N k_B T$ term, to compensate for the missing temperature contribution to the internal energy.

FIG. 8. Internal energy of the composite systems as a function of the number of atoms in the nanolayer. This graph shows nanolayers with initial (100) surface in the Ward potential. Data was normalized to the interface area and a function $h = 3.35\,\mathrm{eV} \cdot N/A$ was subtracted from $U/A$ to exaggerate the difference in slopes for visualization purposes. The blue circles show the results from Figure 7a in the main paper. The yellow diamonds show the results of an MD simulation where the layer atoms are first held in position, followed by a normal MD simulation. Except where marked, the nanolayers stayed crystalline in these simulations. Red squares show a simple molecular statics calculation plus a $2 \times \frac{3}{2}N k_B T$ term, to compensate for the missing temperature contribution to the internal energy.



\end{document}